\documentstyle[twocolumn,prl,aps,epsfig,amssymb,rotate]{revtex}

\newcommand{\beq}{\begin{equation}}
\newcommand{\eeq}{\end{equation}}
\newcommand{\ba}{\begin{array}}
\newcommand{\bea}{\begin{eqnarray}}
\newcommand{\ea}{\end{array}}
\newcommand{\eea}{\end{eqnarray}}

\newcommand\comment[1]{ \hbox{[{\it Comment suppressed here.}\/]} }
\newcommand\hide[1]{}

\newcommand{\skipover}[1]{}

\newcommand{\Tr}{\hbox{Tr}}

\newcommand{\bp}{{\bf p}}
\newcommand{\bq}{{\bf q}}
\newcommand{\bk}{{\bf k}}

\begin{document}                                                
\title{Thermalization of Quantum Fields 
from Time-Reversal Invariant Evolution Equations}
\author{J\"urgen Berges and J\"urgen Cox}
\bigskip
\address{
Center for Theoretical Physics,
Massachusetts Institute of Technology, Cambridge, MA 02139}
%\date{MIT-CTP-2988}
\maketitle
\begin{abstract} 
%************
We study the time evolution of correlation functions
in closed quantum systems for nonequilibrium ensembles 
of initial conditions. For a scalar quantum field theory   
we show that generic time-reversal invariant evolutions 
approach equilibrium at large times. 
The calculation provides a first principles justification 
of Boltzmann's conjecture that the large-time behavior
of isolated macroscopic systems can be described by
thermal ensemble averages. 
%************
\end{abstract}
\begin{narrowtext}

\vspace*{-1.5cm}
\section{Introduction}

\vspace*{-.2cm}
One of the fundamental questions in statistical physics 
concerns the observation of macroscopic irreversible
or dissipative processes arising from microscopic reversible
dynamics. A classical approach to this question 
employs the separation of a system into
observed and unobserved or environmental degrees of freedom and
an averaging procedure over the latter \cite{zwanzig}. The reduced system
may lose the detailed information about initial
nonequilibrium conditions --- a necessary condition to 
reach thermal equilibrium. In contrast, in a closed system with
a unitary time evolution no information is lost and it 
cannot exhibit dissipation or reach thermal equilibrium
at a fundamental level. 

In this Letter we study the
time evolution of closed quantum systems:
for a real, scalar quantum field theory 
we show that generic time-reversal invariant evolutions of nonequilibrium
quantum fields approach thermal equilibrium.
We compute time-reversal invariant 
evolution equations for correlation functions
$\Tr \left\{ \rho(t_0)\, \Phi(t_1)\ldots \Phi(t_n) \right\}$,
i.e.\ expectation values of products of Heisenberg 
fields $\Phi$ for given initial density matrix $\rho(t_0)$.
We explicitly calculate the large-time limit numerically  
for a set of different initial conditions with given average 
energy density. We observe that the correlation functions
approach the thermal distribution asymptotically
\vspace*{-0.1cm}
\bea \!\!\!\!
\lim\limits_{\stackrel{\,\,\,\,\,|t_0-t_i|\to \infty}
{\,\,\,\,i=1,\ldots,n}}
\!\!\!\!\!\!
\Tr \left\{ \rho(t_0) \Phi(t_1)\!\ldots\!\Phi(t_n) \right\}
\!\sim \!\Tr \left\{e^{-\beta H} \Phi(t_1)\!\ldots\!\Phi(t_n) \right\}
\nonumber
\eea
without reaching equilibrium at accessible finite times. Typical
evolutions of correlation
functions for spatially homogeneous fields 
in 1+1 dimensions are shown in 
\mbox{fig.\ 2}. 
Our calculation provides for a quantum field theory 
a first principles justification of 
Boltzmann's conjecture: the large-time behavior of 
isolated macroscopic systems can effectively be described by 
equilibrium ensemble averages.
Apart from its fundamental meaning the treatment of
closed quantum systems is of considerable practical importance. 
In many situations a clear separation into reduced system and environment
is not obvious. This applies in particular to strongly coupled
theories or systems far away from equilibrium where a strict
separation of scales often does not exist. The nonequilibrium description
of many-body quantum systems or
quantum field theories has most diverse applications ranging from 
mesoscopic quantum devices to relativistic heavy-ion collision 
experiments. 

Nonequilibrium quantum field theory has attracted much interest
in recent years. Nonperturbative approaches based on mean field 
or large-$N$ approximations as in \cite{mfcite,CHKMP,Bo} or the 
use of evolution equations for generating functionals \cite{LC}
can give first principle insights into the dynamics of
quantum fields. However, the large-time behavior 
of nonequilibrium quantum field theory 
posed an unresolved problem \cite{CHKMP,Bo,LC}. 
We study the time evolution of correlation functions 
in a mean field approximation,  
and in the Born collision approximation first suggested
by Kadanoff and Baym for nonrelativistic 
real-time thermal Green's functions \cite{KB}. 
Both descriptions lead to causal and manifestly time-reversal invariant 
evolution equations. While the mean field approximation fails to 
describe the long-time behavior \cite{mfcite,CHKMP,Bo,LC,AS}, 
we show that thermalization  
can be described in the collision approximation without further
assumptions.

\section{Method: loop expansion of the $2PI$ effective action} 

\vspace*{-.2cm}
We investigate a quantum field theory for a real, scalar field 
$\varphi$ with classical action 
\bea
\label{classical}
S[\varphi]\!=\!\!\int\!\! {\rm d}^{d+1}x\!
	\left(\frac{1}{2}\partial^\mu\varphi(x)\partial_\mu\varphi(x)  
	\!-\!\frac{m^2}{2}\varphi^2\!(x) 
	\!-\!\frac{\lambda}{4!}\varphi^4\!(x)\! \right) \!\!\!
\eea
The time evolution of $n$-point correlation
functions for a given initial density matrix $\rho(t_0)$ can be obtained
from the generating functional     
\bea
Z_\rho[J,K]&=&\Tr \left\{ \rho(t_0) \, T_c \exp \frac{i}{\hbar}\left(  
\int_{{\cal C}, x} J(x) \Phi(x) \right.\right. \nonumber\\
&& + \left.\left. \frac{1}{2} \int_{{\cal C}, x y}  
K(x,y) \Phi(x) \Phi(y) \right) \right\} 
\label{Z}
\eea
by taking derivatives of $Z$ with respect to the external sources $J$
or $K$. In our notation  $\int_{{\cal C},x} \equiv  
\int {\rm d}^d {\bf x} \int_{\cal C}  {\rm d}x_0$ and 
$T_c$ denotes time ordering along a closed time contour $\cal C$
starting at $t_0$ and chosen to support $J$, $K$ at times of
interest \cite{SK}.
The introduction of the external bilocal source term
in (\ref{Z}) is used to construct the generating functional 
for $2PI$ Green's functions \cite{CJT}.
The evaluation of the $2PI$ generating functional 
for a closed time path has been 
presented by Chou et al.\ \cite{Chou} and by Calzetta \& Hu \cite{CH},
and we refer the reader to their work for details.
The path integral
representation for $Z_\rho[J,K]$ \cite{CH,Chou} 
\bea\lefteqn{
Z_\rho[J,K] \equiv Z[J,K;\alpha_1,\alpha_2,\alpha_3,\alpha_4,\ldots]=
\int\! {\cal D}\varphi 
\exp \frac{i}{\hbar}\Big(S[\varphi]} \nonumber\\ 
&+&\!\int_{{\cal C}, x} 
\!\!\!\!\!\!\left(J(x)\!+\!\alpha_1(x)\right)\! \varphi(x)
\!+\! \frac{1}{2}\! \int_{{\cal C}, x y} \!\!\!\!\!\!\!\!\!
\left(K(x,y)\!+\!\alpha_2(x,y)\right)\! \varphi(x) \varphi(y)\nonumber \\
&+& \frac{1}{3!}
\int_{{\cal C}, x y z} \!\!\!\!\!\!\!
\alpha_3(x,y,z) \varphi(x) \varphi(y) \varphi(z) 
+ \frac{1}{4!} \int_{{\cal C}, x y z w}  \!\!\!\!\!\!\!\!\!
\alpha_4(x,y,z,w) \nonumber\\
&&\varphi(x) \varphi(y) \varphi(z) \varphi(w)
 + \ldots \Big)
\label{Zpath}
\eea
exploits the fact that the information contained in the 
density matrix $\rho(t_0)$ can alternatively be described by
specifying all initial correlations. Here the source terms 
$\alpha_i$ are non-vanishing at initial times $t=t_0$ only. (An explicit
representation of the sources in terms of the density matrix
is given in \cite{Chou}, section 6.) The
$2PI$ effective action is 
defined as the Legendre transform of 
$Z[J,K;\alpha_1,\alpha_2,\alpha_3,\alpha_4,,\ldots]$ with 
respect to $J$ and $K$ \cite{CJT}
\bea
\label{effaction}
\lefteqn{\Gamma[\phi,G;\alpha_3,\alpha_4,\ldots]
=-i \hbar \ln Z[J,K;\alpha_3,\alpha_4,\ldots] } \\
&-&\int_{{\cal C}, x} \!\!\! \phi(x)J(x) 
-\frac{1}{2}\int_{{\cal C}, x y} \!\! 
\Big(\phi(x)\phi(y)+\hbar G(x,y)\Big)K(x,y) \, . \nonumber 
\eea
Here we have absorbed $\alpha_1$ into $J$ and 
$\alpha_2$ into $K$. The effective
action (\ref{effaction}) is
parametrized by the macroscopic field
$\phi(x)\equiv -i \hbar \delta (\ln Z)/\delta J(x) =
\langle \varphi(x) \rangle$ and the exact connected propagator
$\hbar G(x,y)=\langle \varphi(x)\varphi(y) \rangle -    
\langle \varphi(x) \rangle\langle \varphi(y) \rangle$
in the presence of the initial-time source terms $\alpha_3,\alpha_4$ etc. 
Solutions for $\phi$ and $G$ require 
\bea
\frac{\delta \Gamma[\phi,G;\alpha_3,\alpha_4,\ldots]}
{\delta \phi(x)}&=&-J(x) - \int_{{\cal C}, x} \!\! K(x,y) \phi(y) \, , 
\nonumber \\ 
\frac{\delta \Gamma[\phi,G;\alpha_3,\alpha_4,\ldots]}{\delta G(x,y)}
&=&-\frac{1}{2} \hbar K(x,y) \, .
\label{station}
\eea 
where the RHS is zero for vanishing external sources $J$, $K$ ($t>t_0$). 
Eq.\ (\ref{station}) yields time evolution equations for the macroscopic
fields $\phi$ and $G$. We note that these equations are obtained 
from an action functional by a variational principle which guarantees a 
unitary time evolution.   
We restrict the discussion to a quartic initial density matrix,
i.e.\ $\alpha_i=0$ for $i\ge 5$, which is no approximation but
constrains the initial state. We approximate 
the effective action (\ref{effaction}) by a series 
expansion in orders of $\hbar$ or, equivalently, an expansion
in the number of loops for two-particle irreducible 
graphs \cite{CJT,CH,Chou}.  
For simplicity we consider in this section the three-loop result 
for the effective action for a vanishing field
expectation value $\phi$. Deviations from $\phi=0$ can be included
in a straightforward way and results are described below. 
For the $Z_2$-symmetric theory this implies $\alpha_3\equiv 0$
and the effective action $\Gamma[G;\alpha_4]\equiv \Gamma[0,G;0,\alpha_4]$
reads to order $\hbar^3$   
\begin{eqnarray}
\lefteqn{
\Gamma[G;\alpha_4]\!=\!\!
\frac{i\hbar}{2}\!\int_{{\cal C}, x}\!\!\!\!\!
[\ln G^{-1}](x,x)\!-\!\frac{\hbar}{2}\!\int_{{\cal C}, x y} \!\!\!\!\!\!\!\!
\delta_{x,y}(\square_x+m^2) G(x,y)}\nonumber\\
&-&\frac{\hbar^2}{8}\int_{{\cal C}, x y z w} \hspace*{-0.5cm}
(\lambda \delta_{x,y}\delta_{x,z}\delta_{x,w} -\alpha_4(x,y,z,w)) 
G(x,y) G(z,w)
\nonumber\\
&-& \frac{\hbar^3}{48}
\int_{{\cal C}, x y z w} \hspace*{-0.5cm}
(\lambda \delta_{x,y}\delta_{x,z}\delta_{x,w} -\alpha_4(x,y,z,w)) 
L(x,y,z,w)
\label{effectiveaction}
\end{eqnarray}
with the four-point function
\bea
\label{propver}
\lefteqn{L(x,y,z,w)=-i
\int_{{\cal C}, x' y' z' w'} \hspace*{-0.5cm} 
(\lambda \delta_{x',y'}\delta_{x',z'}\delta_{x',w'} }\nonumber\\
&&-\alpha_4(x',y',z',w')) 
G(x,x') G(y,y') G(z,z') G(w,w') \, .
\eea

\vspace*{-.2cm}
\section{Nonequilibrium time-reversal invariant dynamics}
\label{evolutioneqs}

\vspace*{-.2cm}
The time evolution equation for the time ordered propagator 
\beq
\label{timeordered}
G(x,y)= G_>(x,y)\Theta(x_0-y_0)+G_>(y,x)\Theta(y_0-x_0)
\eeq
is obtained from (\ref{station}) 
and the equal-time commutation relation
$i\frac{d}{dx_0}\left.\left(G_>(x,y)-G_>(y,x)\right)\right|_{x_0=y_0}\!\!\!=
\delta_{x,y}$ for vanishing external source $K$
\bea
\label{GEvol}
\square_x G_>(x,y)&=&-\left(
m^2 +\frac{\lambda \hbar}{2} G_>(x,x) \right)
G_>(x,y) \nonumber \\
&&-\frac{\lambda \hbar^2}{6} L(x,y,y,y)  \,\, 
\eea
with the four-point function 
\bea
\lefteqn{
\label{vertexcausal}
L(x,y,y,y) = - i\lambda \int {\rm d}^d {\vec z} 
\,\bigg\{\int_0^{y_0}\!\! {\rm d}z_0 
 G_>(y,z) G_>^3(x,z) }\nonumber\\
&&
+ \int_{y_0}^{x_0}\!\! {\rm d}z_0
 G_>(z,y) G_>^3(x,z)  
- \int_{0}^{x_0}\!\! {\rm d}z_0  
 G_>(z,y) G_>^3(z,x)  \bigg\} \nonumber\\
&&+ \, L_0({\bf x},{\bf y},{\bf y},{\bf y}) \, .
\eea
Here $L_0$ is the initial-time four-point function which is determined by
(\ref{propver}) for given initial $\alpha_4$ and $G$. Note 
that $G_>^*(x,y)=G_>(y,x)$. The evolution for $G_>(x,y)$ is time-reflection
invariant and causal since the ``memory integral'' 
in (\ref{vertexcausal}) only depends on times smaller than max($x_0,y_0$).

The evolution equation 
for $G_>$ is a nonlinear, integral-differential equation 
which can be solved numerically. We use a standard lattice
discretization for a spatial volume $V$ with periodic boundary 
conditions and study the large volume limit to remove finite
size effects. The time discretization respects time-reversal 
invariance. Numerical results are calculated for 1+1 dimensions.
All quantities will be expressed in units of appropriate
powers of $m$ with $\hbar\equiv 1$.
We consider spatially homogeneous fields such that
$G_>(x,y)=G_>(x_0,y_0;{\bf x}-{\bf y})$ and Fourier transform
with respect to the spatial variables.
The initial conditions are specified in terms of the 
momentum modes of the propagator $G_>(0,0;{\bf p})$, 
its first derivative with respect to time $\dot{G}$, and the 
four-point function $L_0$ at initial time.

\vspace*{-.5cm}
\subsection{Two-loop (mean field) approximation}
\label{meanfieldsection}

\vspace*{-.3cm}
The solution of (\ref{GEvol}) for the propagator to order $\hbar^0$ 
corresponds
to the free field solution and the propagator modes $G_>(t,0;\bp)$ 
oscillate with frequency $\sqrt{{\bf p}^2+m^2}/2 \pi$.  
The two-loop contribution to the  
effective action (\ref{effectiveaction}) adds a time dependent mass shift 
$\Delta m(t)= \frac{\lambda \hbar}{2} \int\!\! \frac{d^d \bq}{(2 \pi)^d}\,
G_>(t,t;\bq)$ to the free field evolution equation.
As a consequence the equation (\ref{GEvol}) for $G$ becomes nonlinear. 
The mass shift term in the evolution equation is
the same for all Fourier modes $G_>(t,t';{\bf p})$
and each mode propagates collisionlessly with a time dependent
effective mass.
\mbox{Fig.\ 1} shows a typical time evolution of the mass shift
$\Delta m(t)$ to order
$\hbar$.  As initial conditions
we use the mean field thermal solution for $G(0,0;\bp)$ with 
inverse temperature $\beta=0.5$ and $\dot{G}(0,0;\bp)$ 
chosen to deviate from the 
thermal solution. For $\Delta m(t)$ 
one observes an initial effective damping of oscillations in the mean
field approximation termed dephasing \cite{mfcite}.
The damping is more efficient in higher dimensions \cite{mfcite,CHKMP,Bo,LC}.
Since the evolution equation is time-reversal invariant
dissipation is absent and the effective damping is due to 
the superposition of oscillatory functions with continuous 
frequency spectrum. This absence of dissipation can be easily
demonstrated for finite systems with a discrete frequency spectrum.
In this case the time-reversal invariant evolution 
equation, which is local in time to order $\hbar$, leads to characteristic 
recurrence times after which the effective damping is lost.
Since we use a lattice discretization for a 
spatial volume $V$ we have a discrete frequency spectrum for
finite volumes. 
In 1+1 dimensions we explicitly verify that the observed
recurrence times for $\Delta m(t)$ scale with $V$ or the
number of degrees of freedom to infinity. 

The lower curve in \mbox{fig.\ 1} shows a
typical evolution of the macroscopic field $\phi(t)$ in the
mean field approximation with nonzero initial $\phi$.
The evolution of the field, but also the evolution of 
Fourier modes of correlation functions, are typically undamped
in 1+1 dimensions. 
The mean field approximation does not describe the 
approach to a thermal equilibrium distribution 
\cite{mfcite,CHKMP,Bo,LC,AS}. 
\vspace*{-0.6cm}
\begin{figure}
\begin{center}
\rotate[l]{\epsfig{file=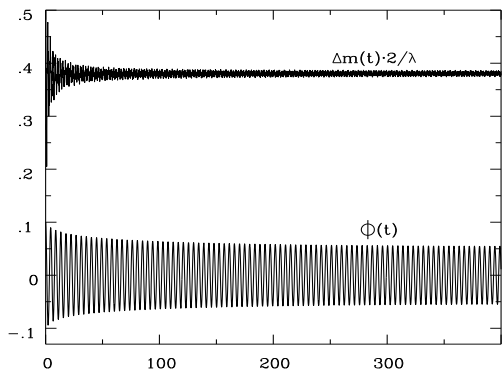,width=2.5in,height=4.1in}}
\end{center} \label{gmf2}
\vspace*{-1.5cm}
\caption{Mass shift (upper curve) and macroscopic field (lower curve)
in mean field approximation.}
\end{figure}
%

%\vspace*{-0.3cm}
\subsection{Three-loop (collision) approximation: 
\mbox{thermalization}}

\vspace*{-.3cm}
A crucial improvement for the description of the long-time
behavior comes from the three-loop contribution to the
effective action (\ref{effectiveaction}). 
The resulting effective four-point 
%\noindent
function (\ref{vertexcausal}) in the 
evolution equation for $G_>$ introduces 
scattering. 
In \mbox{fig.\ 2} we show the
time dependence of the equal-time propagator $G_>(t,t;\bp)$
for three Fourier modes $|\bp|=0,3,5$ and three different 
initial conditions with the same energy density.
For the solid line the initial conditions are close
to a mean field thermal solution with $\beta=0.1$, the initial mode 
distribution for the dashed and the dashed-dotted lines deviate 
more and more from thermal equilibrium.  
It is striking to observe that propagator modes 
with very different initial values but with the same momentum $\bp$ 
approach the same large-time value.
The asymptotic behavior of the two-point 
function modes are uniquely determined by the initial 
energy density
which is characteristic for thermal equilibrium.
\vspace*{-0.4cm}
\begin{figure}
\begin{center}
\rotate[l]{\epsfig{file=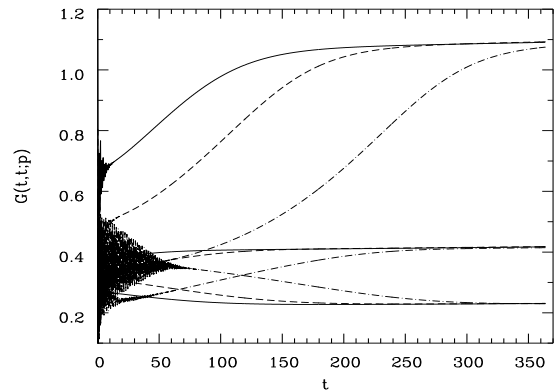,width=2.5in,height=4.1in}}
\end{center} \label{unidyn}
\vspace*{-1.3cm}
\caption{Momentum modes of the two-point function 
for different initial conditions with same energy.}
\end{figure}
We explicitly verify that energy is conserved
and that the higher $n$-point functions approach asymptotic
values uniquely determined by the energy.
Note that all higher correlation functions are expressed
in terms of $G$ and their effective 
thermalization results from the behavior of the
two-point function. We find that small deviations 
from $\phi=0$ approach zero in the long-time limit for the
initial conditions considered in \mbox{fig.\ 2}.    

The evolution of correlation function modes exhibits an effective damping of 
oscillations. The decrease of the maximum amplitude quickly approaches an 
exponential behavior, but the oscillations never damp out
completely. This can be seen from the logarithmic plot of the propagator
zero mode $G_>(t,t';0)$ in \mbox{fig.\ 3}. 
The upper curve shows $|G_>(t,0;0)|$ and 
exhibits a strong effective reduction
of correlations of the field at time $t$ with the initial field.
The lower curve presents the equal-time propagator $G_>(t,t;0)$,  
subtracted by its time average over $20$ time steps 
$\overline{G}_t$ and limited below to make it 
suitable for a logarithmic plot. 
\vspace*{-0.4cm}
\begin{figure}
\begin{center}
\rotate[l]{\epsfig{file=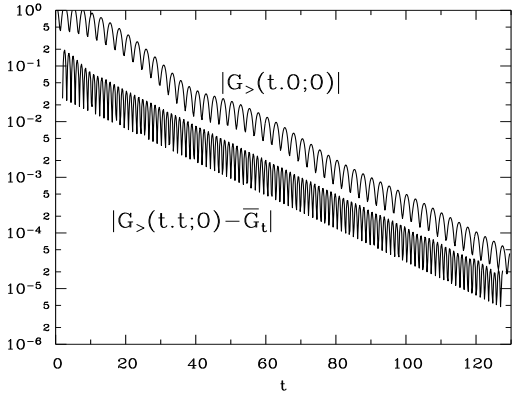,width=2.5in,height=4.1in}}
\end{center} \label{g0tb}
\vspace*{-1.5cm}
\caption{Logarithmic plot for the propagator
zero mode.}
\end{figure}
We emphasize that time-reversal invariance is manifest, i.e.\ 
for each evolution towards the thermal solution there exists the 
reversed evolution away from equilibrium. However, if one follows 
these reversed evolutions generically the system approaches equilibrium 
again after large enough times. We note that
recurrence times can be infinite in the collision approximation
even for small volumes $V$ with discrete Fourier
momentum modes. This is possible because the evolution equation 
for the ensemble average (\ref{GEvol}),(\ref{vertexcausal})  
is nonlocal in time. The memory effects quickly lead to 
contributions which can
be described by the superposition of oscillatory functions with
a continuous frequency spectrum. For the initial conditions
employed in \mbox{figs.\ 2} simulations for rather small 
volumes of order tens
the correlation length are found to describe the large-time 
behavior well.

The efficient damping observed in \mbox{figs.\ 2}
is due to a large coupling constant $\lambda=10$. 
The damping time $\tau$, defined as the
time for which the envelope amplitude of $G_>(t,t;0)$
is reduced by a factor $e$, is presented in \mbox{fig.\ 4} as
a function of $\lambda$. We find that the critical coupling
for which the amplitudes are undamped is zero. The qualitative 
thermalization behavior does 
not differ in the weak and the strong coupling regime
in the present approximation, which neglects higher-order scattering 
effects. These are relevant, in particular, for the computation
of the bulk viscosity or the description
of critical phenomena \cite{bulk,CH}.  

Of course, there are examples, like an energy eigenstate as 
initial condition, where thermalization can not occur. 
For few body systems like a set of coupled anharmonic 
oscillators it is known that a transition to non-ergodic behavior
can occur for small but nonzero couplings \cite{Bohigas}. 
For a finite volume $V$
and ultraviolet cutoff for the momentum modes 
our model can be mapped onto a finite set of coupled anharmonic
oscillators. We indeed find that    
for small enough volumes and couplings below a certain critical 
coupling $\lambda_c$ the long-time evolution can be undamped.    
The values for the critical coupling strongly 
depend on the initial conditions and the volume or number of degrees
of freedom.
\vspace*{-0.4cm}
\begin{figure}
\begin{center}
\rotate[l]{\epsfig{file=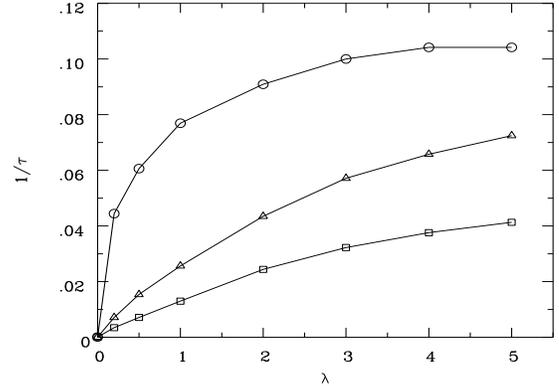,width=2.5in,height=4.1in}}
\end{center} \label{damptime}
\vspace*{-1.3cm}
\caption{Damping time $\!\tau\!$ for propagator modes
with momenta $\!|\bp|\!=\!0,3,5\, \mbox{\rm (from top)}$ 
as a function of the coupling $\!\lambda$.}
\end{figure}

To verify the equivalence of the large-time limit of 
nonequilibrium correlation functions with the description
in terms of a thermal distribution, we directly compute 
equilibrium correlation functions for $\rho=\exp{(-\beta H)}$.
The computation of the quantum-statistical partition
function is done in the standard imaginary-time
formalism, equivalent to evaluating (\ref{Z}) for an
imaginary time path $C$ running from zero to $-i \beta$.
From the stationary condition for the corresponding 
imaginary-time three-loop
effective action one obtains the gap equation for the 
propagator Fourier modes
\bea 
\lefteqn{\!\!\!\!\!\!\!
G_E^{-1}(\omega_l,\bp)= w_l^2+\bp^2\!+m^2\!
+\frac{\lambda \hbar}{2} T \sum\limits_n\! \int\!\! \frac{d^d\bq}{(2 \pi)^d}
G_E(\omega_n,\bq)} \nonumber\\
&-&\frac{\lambda^2 \hbar^2}{6} T^2 \sum\limits_{n,m}
\int \frac{d^d\bq}{(2 \pi)^d}\frac{d^d\bk}{(2 \pi)^d}\,
G_E(\omega_n,\bq) G_E(\omega_m,\bk) \nonumber\\
&&G_E(\omega_l-\omega_n-\omega_m,\bp-\bq-\bk) \, .
\label{img}
\eea
Here $\omega_l=2\pi l/\beta$, $l\! \in\! Z$ and the relation between the
Euclidean propagator $G_E$ and the equal-time propagator $G$ is
given by $\,\, T \sum_l G_E(\omega_l,\bp)=G(t-t;\bp)$.
We solve eq.\ (\ref{img}) by iterating the RHS, starting 
from the mean field solution, until convergence is obtained.
We find that the large-time result for the equal-time correlator from 
the real-time dynamics and from the partition function agrees
in the continuum limit. Their correspondence 
provides a first principles justification of Boltzmann's conjecture.

\vspace*{0.4cm}
J.B.\ thanks E.\ Mottola for discussions
about the $2PI$ effective action and F.\ Cooper,
T.\ Papenbrock and J.\ Verbaarschot for 
helpful conversations. 
Research supported by the DOE under agreement 
DE-FC02-94ER40818 and by the INT at the University of 
Washington where part of this work has been done.   

\end{narrowtext}

\end{document}